# Characterization of hydrogen plasma in a permanent ring magnet based helicon plasma source for negative ion source research

A. Pandey[1,2], Debrup Mukherjee[1,2], M. Bandyopadhyay[1,2,3], Dipshikha Borah[1,2], Himanshu Tyagi[3], Ratnakar Yadav[3], A. Chakraborty[1,3]

[1]Institute for Plasma Research, Bhat, Gujarat, 382428 India

[2]Homi Bhabha National Institute (HBNI), Mumbai, India

[3]ITER-India, Institute for Plasma Research, Gujarat, India

email: mainak@iter-india.org

Abstract

HELicon Experiment for Negative ion source (HELEN-I) with single driver is developed with a focus on the production of negative hydrogen ions. In the Helicon wave heated plasmas, very high plasma densities ($\sim 10^{19} m^{-3}$) can be attained with electron temperatures as low as ~ 1 eV in the downstream region. These conditions favor the production of negative hydrogen ions. In HELEN-I device at IPR, helicon plasma is produced using Hydrogen gas in a diverging magnetic field, created by a permanent ring magnet. RF Power ($P_{RF}$) of 800-1000W at 13.56 MHz frequency is applied to a Nagoya-III antenna to excite m = 1 helicon mode in the plasma. The plasma is confined by a multi-cusp field configuration in the expansion chamber. The transition from inductively coupled mode to Helicon mode is observed near $P_{RF}$ ~ 700W with plasma density ~ $10^{18}$ m$^{-3}$ and electron temperature ~ 5 eV in the driver and ~ 1eV in the expansion volume. Line integrated negative hydrogen ion density is measured in the expansion chamber by employing an Optical Emission Spectroscopy (OES) diagnostic technique using $H_\alpha/H_\beta$ ratio and Laser photo-detachment based Cavity Ring Down spectroscopic (CRDS) diagnostic technique. The measured value of negative hydrogen ion density is in the order of $10^{16}$ m$^{-3}$ at 6 mTorr pressure and does not vary significantly with power in the helicon mode, pressure and downstream axial magnetic field variation. The negative ion density measurements are compared with theoretically estimated values calculated using particle balance method considering different reaction rates responsible for negative hydrogen ion creation and destruction. It is to be noted that at present Caesium (Cs) is not injected in the plasma discharge to enhance $H^-$ ion density.

## I. Introduction

Negative hydrogen ion ($H^-$) source based Neutral Beam Injector (NNBI) is one of the most efficient auxiliary heating and current-drive systems for fusion plasma reactors [1, 2]. A great deal of interest is drawn towards the development of high density negative ion sources for neutral beam applications. Consequently, R&D activities are set in motion in Europe, Japan and India for developing negative hydrogen sources for NBI systems to be used in large magnetic fusion reactors like ITER. [3-6]. At high energies (> 200 keV) the neutralization efficiency of positive hydrogen ion practically goes to zero. Whereas for $H^-$ ion beam, above the beam energy desired for a conventional gas based neutralizer cell in a NBI system, the neutralization efficiency is still 60% [7]. Considering the reactions responsible for $H^-$ ion production and destruction, it is found that low pressure, high density but low temperature hydrogen plasmas are suitable for an efficient negative hydrogen ion source [8, 9, 10, 11]. The production of negative hydrogen ions takes place through two processes: (a) Volume process and (b) Surface process.

In the volume process, $H^-$ ions are created in a two-step collisional reaction:
1$^{st}$ step: $e_{fast} + H_2 \rightarrow H_2^*(v") + e$, (vibrational excitation of ground state by energetic electrons). 2$^{nd}$ step: $H_2^*(v") + e_{slow} \rightarrow H^- + H$ (vibrationally excited molecule form $H^-$ ion through dissociative attachment (DA) process with a low energy electron). Conventionally, a volume $H^-$ source is divided into two operational parts, the "hot driver region" where $H_2$ molecules are vibrationally excited and the "cold extraction region" where dissociative attachment process takes place. Virtual division of the plasma into two different electron temperatures is maintained inside a plasma volume by applying a transverse magnetic filter field [9]. Low electron temperature across the filter field, near extraction region of an ion source is desirable to prevent $H^-$ destruction due to electron detachment reaction [7, 9]. But the filter field also reduces the electron density near the extraction, which is not a desirable condition for the production of $H^-$ ions.

In surface process, H- ion are produced on a low work-function surface due to surface conversion of energetic H atoms ($H_0^*$) or hydrogen ions ($H_n^+$) in the plasma [9]; $H_0^*/H_n^+ \xrightarrow{Low\ work-function\ surface} H^-$. Low work-function surface is normally created by Caesium (Cs) deposition, by injecting Cs vapour into the plasma. Surface process is more efficient than the volume process. However, despite the higher efficiency of surface process, there are some specific issues linked with the uncontrolled behaviour. Recently, Kakati et al. [11] have reported a proof of the principle negative hydrogen ion source based on a novel concept of surface assisted volume process using Caesium coated tungsten dust dispersed in the volume. In this paper, we present a configuration for a Helicon wave heated plasma which has high plasma density and low electron temperature, suitable for the $H^-$ ion source without using the filter field or Cs in the setup.

The above discussion indicates that a plasma source having high plasma density and low plasma temperature is a suitable candidate for a volume $H^-$ ion source. Briefi and Fantz [12] proposed showed the strategy of the approach for using Helicon discharges for ITER reference source parameters and argued that the helicon plasma sources could be used as $H^-$ ion sources for future NBI systems. The permanent ring magnet based helicon source presented in this paper has the potential to fall in this category. In this paper, we focus on the characterization of hydrogen plasma and subsequently include the negative ion measurement diagnostics in the helicon plasma source to measure $H^-$ ion density. The paper is planned as follows: section II contains the details of the HELEN-I experimental setup including different diagnostics. Section III describes different diagnostic techniques used for the $H^-$ ion density measurement. The measurement results are shown in section IV. In section V we discuss and summarize the results.

## II. Experimental Set-up

A schematic of the experimental device is shown in figure 1. The apparatus consists of a 70 mm long glass plasma source of 50 mm diameter and an SS expansion chamber of 100 mm inner diameter and ~300 mm in length. A permanent NdFeB magnet of surface magnetic field ~4.6 kG is placed above the plasma source to provide an axial field for the excitation of Helicon wave in the plasma. The separation between the magnet and the glass chamber is kept in such a way that magnetic field ~ 40 -100 G is present in the glass chamber volume. The field strength in the source is varied by changing the separation between the magnet and the top flange.

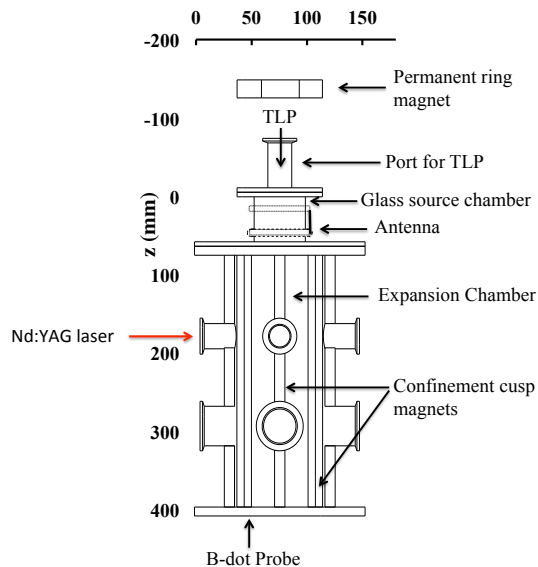

Figure 1. Schematic of HELEN-I set-up.

The expansion chamber has an array of magnets arranged to form a multi cusp field in the chamber. A more detailed description of the experimental apparatus is given in ref. 13. Figure 2a shows the diverging magnetic field in the source and expansion chamber, which depends on the distance of the permanent magnet from the plasma source. Figure 2b shows the cusp field geometry in the expansion chamber forming a field free region of ~ 50 mm diameter.

Hydrogen plasma is produced by applying RF power to the Nagoya III antenna, of 36 mm length, at 13.56 MHz to excite m = ±1 azimuthal helicon mode in the plasma. Optical emission spectroscopy and Laser aided negative ion density diagnostics are used in the set-up as shown in the figure (1) and (3). The light collecting optics is kept at the location z = 19 cm for OES measurement, the Nd:YAG laser beam enters into the expansion chamber through the diagnostic port as shown in the figure. The details of the diagnostics are given in the next section.

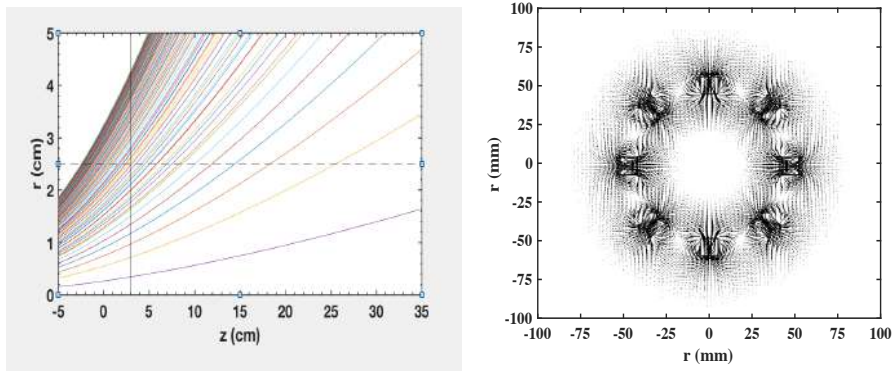

Figure 2. Simulated magnetic field lines in the system are shown: a) diverging axial field lines from the axial field magnet, b) multi cusp field lines from the confinement magnets in the expansion chamber.

### III. Diagnostic Techniques

1. *Electrical probes*:

Two axially movable Triple Langmuir probes (TLPs) are used for measuring the plasma density at different locations in the plasma chamber. One of the probes is installed at the top for the axial measurements in the source and the other probe is introduced through the bottom flange, which is L-shaped and used for axial as well as radial measurements in the expansion chamber. The probes are made from a 0.3 mm diameter tungsten wire and have three identical tips of length = 5 mm each. The axis of rotation of the L-shaped probe is kept slightly off-centred and the radial profile is obtained by the rotation of the probe shaft. Triple probe is less prone to RF noise compared to an uncompensated single Langmuir probe. It gives a relatively reliable estimate of plasma density but the temperature value may be slightly overestimated [14]. A B-dot probe [15], made from 0.5 mm thick enamelled copper wire, with a loop diameter of ~ 2 mm is also used to detect the signature of the helicon wave in the expansion chamber. The locations of the probes are shown in figure 1.

2. *Optical Emission Spectroscopy (OES)*

A non-invasive diagnostic technique based on OES method has been used to measure the negative ion density. Fantz and Wunderlich [7] have demonstrated the OES based technique to measure the negative hydrogen ion density using $H_\alpha, H_\beta$ and $H_\gamma$ Balmer lines. OES is a widely used plasma diagnostic technique for the measurement of plasma parameters such as electron density, electron temperature [6, 7] etc. This technique measures the wavelength resolved line of sight (LOS) averaged value of the atomic or molecular line emission intensity from the plasma in the visible range with the help of a standard spectrometer. Light from the plasma is focused at one end of an optical fibre using a lens and is transmitted to the spectrometer entrance slit. The light is then allowed to fall on the detector through a dispersive medium and wavelength resolved intensity profile is recorded. In hydrogen plasma, $H_\alpha, H_\beta, H_\gamma$ and $H_\delta$ Balmer series lines from atomic transition and Fulcher band from the molecular transitions are dominant.

In case of low pressure and low temperature hydrogen plasma, the processes contributing to the population of excited state of atomic hydrogen leading to $H_\alpha$ emission are listed in table 1. The population coefficients for these processes can be calculated as a function of plasma parameters from CR model. The individual contribution of each species to the Balmer lines for different plasma parameters have been discussed in [7]. Out of all the processes listed in table 1, the contribution from the mutual neutralization $H^-$ ions to emit $H_\alpha$ emission having rate coefficient, $X_{H_\alpha}^{eff,H^-}$, dominates for the electron density range $10^{16} - 10^{21}\ m^{-3}$. The effective rate coefficients for mutual neutralization, $X_{H_{\alpha,\beta,...}}^{eff,H^-}$ are independent of electron temperature as the process involves interaction of ions [7] only. The line ratios $H_\alpha/H_\beta$, from all the processes, listed in table-1 show weak dependency on plasma density. Only the process "mutual neutralization" shows strong dependence on electron density for the line ratio $H_\alpha/H_\beta$. Due to this dependency of Balmer lines on $H^-$ density, the value of the line ratio $H_\alpha/H_\beta$ can be used for the measurement of negative ions. These assumptions hold good only for low pressure plasmas (<1 Pa or 7.5 mTorr). At higher pressures, other factors also become important like self-absorption of Lyman lines and dissociative recombination of $H^{3+}$, adding another channel of radiative process apart from the processes listed in table 1. Thus, at higher pressures, this analysis might give erratic results [7].

Effective excitation and mutual neutralization are the two dominant processes for the quantum number $p = 3$ and $p = 4$ transition and responsible for these two lines. Hence, only these two processes are taken into account while calculating the line ratio $H_\alpha/H_\beta$. Thus the line ratio is given by:

$$\frac{H_\alpha}{H_\beta} = \frac{n_H X^{eff,H}_{H_\alpha} + n_{H^-} X^{eff,H^-}_{H_\alpha}}{n_H X^{eff,H}_{H_\beta} + n_{H^-} X^{eff,H^-}_{H_\beta}} \qquad (1)$$

Fantz and Wunderlich [7] have shown that the contribution of mutual neutralization on the population of quantum numbers p > 4 and thus $H_\gamma$, $H_\delta$, etc, are negligible. But, to obtain the $H^-$ density, we need to know the atomic hydrogen population. In order to determine the density of atomic hydrogen, absolute measurement of $H_\gamma$ line emission can be used. The line emission $H_\gamma$ is given by:

$$H_\gamma = n_H n_e X^{eff,H}_{H_\gamma}(T_e, n_e) \qquad (2)$$

Using eq.(1) and eq.(2), a relation can be obtained for known value of electron temperature, $T_e$ and electron density, $n_e$

$$n_{H^-} = \frac{H_\gamma}{n_e} C_1 \left(1 - \frac{H_\alpha}{H_\beta}\frac{1}{C_2}\right)\left(1 - \frac{H_\alpha}{H_\beta}\frac{1}{C_3}\right)^{-1} \qquad (3)$$

Where, $C_1 = \frac{X^{eff,H}_{H_\alpha}}{X^{eff,H}_{H_\gamma}} \frac{1}{X^{eff,H^-}_{H_\alpha}}$; $\quad C_2 = \frac{X^{eff,H}_{H_\alpha}}{X^{eff,H}_{H_\beta}}$; $\quad C_3 = \frac{X^{eff,H^-}_{H_\alpha}}{X^{eff,H^-}_{H_\beta}}$;

Table 1: Processes contributing to excited state of atomic hydrogen

| Processes |
|---|
| 1. Recombination: $H^+ + e \rightarrow H(p)$ |
| 2. Dissociative recombination: $H_2^+ + e \rightarrow H(p) + H$ |
| 3. Effective excitation: $H + e \rightarrow H(p) + e$ |
| 4. Mutual neutralization: $H^- + H^+ \rightarrow H(p) + H$ |
| 5. Dissociative excitation: $H_2 + e \rightarrow H(p) + H + e$ |

The factor $C_1$ represents the ratio of effective excitation rate coefficient for $H_\alpha$ and $H_\gamma$ divided by the effective mutual neutralization rate coefficient for $H_\alpha$ emission. The factor $C_2$ can be interpreted as the line ratio $H_\alpha/H_\beta$ when the excitation takes place from atomic hydrogen only. The factor $C_3$ represents the line ratio $H_\alpha/H_\beta$ when the excitation takes place from negative hydrogen ions only. These factors can be calculated using CR model for known plasma parameters [16].

### 3. Cavity Ring Down Spectroscopy (CRDS)

The CRDS technique, developed by O'Keefe and Deacon, is based upon the measurement of the rate of absorption of a light pulse confined within a closed optical cavity [17]. It is a very sensitive diagnostic technique developed to measure the density of particles present in a trace amount [18, 19], e.g. negative hydrogen ion in the present case. Recently, Agnello et al. have shown that CRDS can be used for measuring the $H^-$ ion density in a hydrogen helicon plasma where the $H^-$ ion population is small compared to the $H^+$ ions [20].

When a photon of energy higher than the electron affinity of H atoms (0.75 eV) enters into a cavity formed by two highly reflecting mirrors at the two ends of the cavity volume filled with $H^-$ ions, loosely bound electron in $H^-$ ion is detached by the laser photon due to laser photo-detachment reaction and the corresponding photons are considered as absorbed. In the present experiment, beam from a Nd:YAG (Innolas make) infrared laser of 1064 nm wavelength is fed into the plasma in the expansion chamber as shown in figure 3 through a 99.99% reflecting mirror, m1 (attached to left side port at z = 19 cm from the endplate). Only 0.01% of the photons go into the cavity. The port on the opposite end of the laser entrance attached to a long (900 mm) trapping cavity also has a mirror, m2, identical to m1. This arrangement of two highly reflecting mirrors forms a cavity where the laser is trapped to make multiple reflections and so multiple pass through the plasma volume. Due to the cavity configuration by two mirrors, remaining laser photons bounce back and forth multiple times between the two mirrors. In each of these trips through the plasma medium, a fraction of total number of photons gets absorbed with time. As a result, a temporally decaying laser intensity profile is recorded by a detector kept just outside the cavity. The decay time constant, known as ring down time (RDT) is used to calculate the density of $H^-$ ions inside the plasma. Just outside the cavity, a photodetector is placed to receive the photons transmitted through m2. The laser beam is 2 mm in diameter with a pulse width of ~ 6 ns. The maximum beam energy can be as high as 180 mJ. The energy of 1064 nm photon is 1.2 eV sufficiently higher than the electron affinity (0.76 eV) of H atom to form $H^-$ ions and able to detach the loosely bound electron.

In vacuum without plasma or $H^-$ ions, the exponential decay is relatively slow ($T_1$) due to unavailability of absorbing medium for that wavelength photon. In this condition, the laser is scattered less. However, in the case of a medium (plasma with H⁻ ions) the interaction is stronger and thus the decay is faster ($T_2$). Thus, we measure the time constant of the signal to compute the negative hydrogen ion density from the relation,

$$n_{H^-} = \frac{L}{dc\sigma}\left(\frac{1}{T_2} - \frac{1}{T_1}\right) \quad (4)$$

where, *L = 120 cm* is the total length of the cavity, *d = 10 cm* is the length of absorbing medium traversed by the photon, $\sigma = 3.5 \times 10^{-17}$ *cm²* is the photo-detachment cross-section or photon absorption cross-section [21]. A detailed CRDS development description and a parametric study of negative hydrogen ion density measurement in HELEN using CRDS are done by Debrup et al. [22, 23].

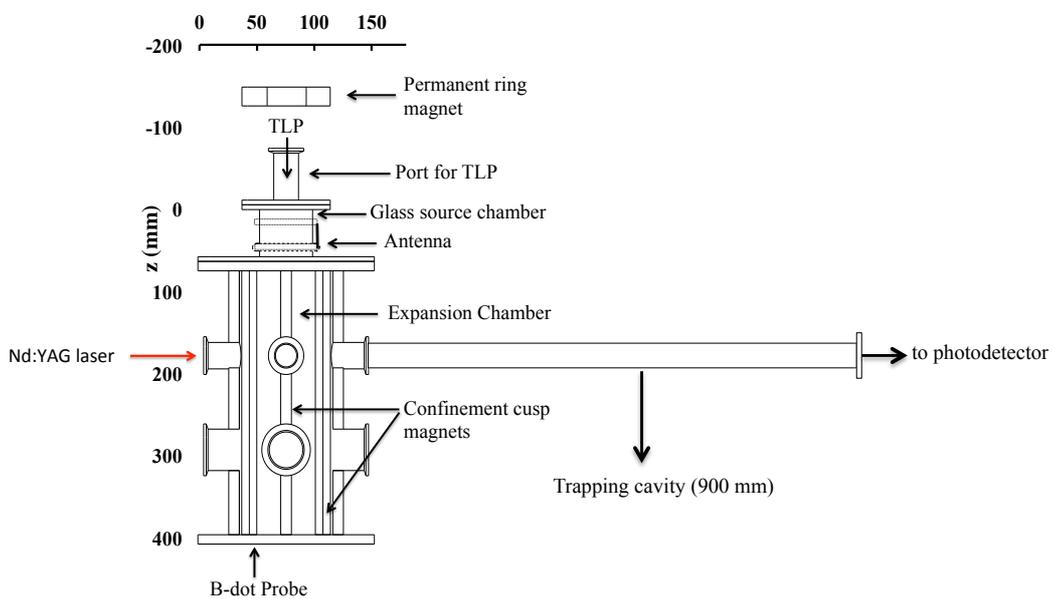

Figure 3. HELEN-I experimental set-up attached with ~1m length CRDS cavity.

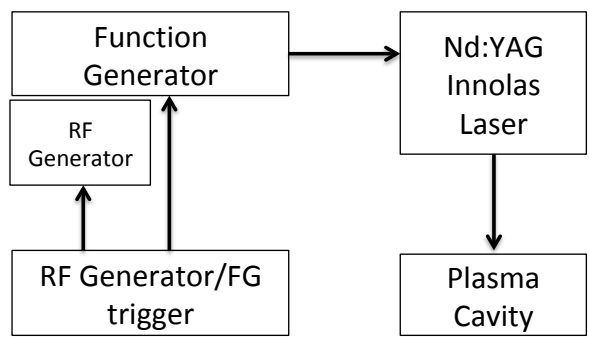

Figure 4. Schematic of the laser input system for the CRDS experiment.

A schematic of the laser triggering system is shown in figure 4. The input trigger for the RF Generator also serves as the input trigger for the function generator, which generates the input signal for triggering the laser after a delay of 50 ms.

Both OES and CRDS measure line integrated $H^-$ ion density, but due to axial magnetic field and associated cusp magnetic field structure in the present Helicon source, a non-uniform profile of $H^-$ ion density is expected. So, a local $H^-$ ion measurement is needed to study the profile. To measure local $H^-$ion density to generate its profile, the Laser Photo-Detachment (LPD) diagnostic is under development and will be reported in future.

## IV. Results and Discussion

### a. Measurement of plasma parameters

Typical signatures of helicon mode operation are monitored through a sudden jump in optical light emission and in ion saturation current by the Langmuir probe. Fig.5 shows the ion saturation current jumps in hydrogen plasma with a Nagoya antenna and RF power greater than 400W. This density jump is not obtained with a loop antenna because of low induction field inside the plasma [24].

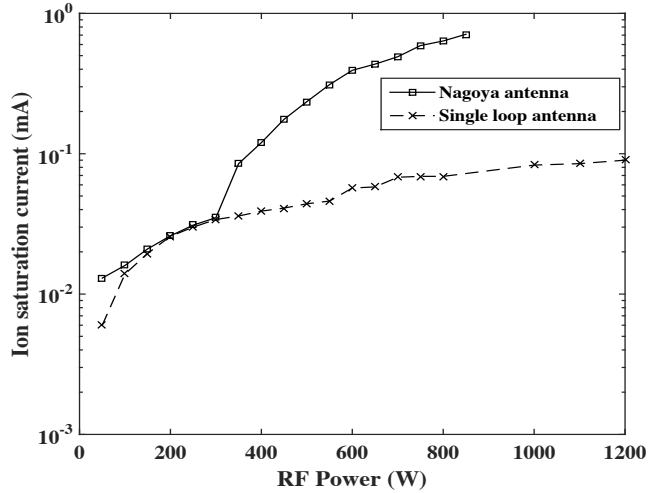

Figure 5 Ion saturation jump in hydrogen plasma while using Nagoya antenna and power is around 400W or above.

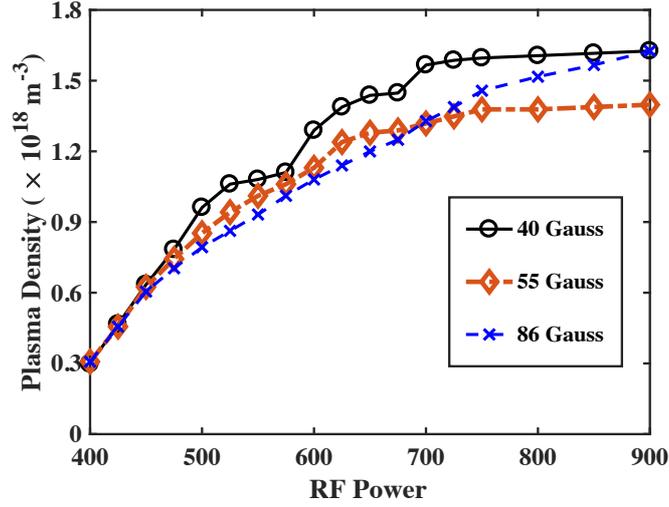

Figure 6. Variation of ion saturation current with increasing RF power at 6-mTorr pressure. A clear jump from capacitive to inductive and subsequently to the helicon mode is seen from the figure. The axial field value shown is the maximum field in the source at z = 0.

It can be seen from figure 6 that a stable helicon plasma mode exists after 700 W. Further diagnosis is done at 800 W, where the hydrogen plasma is helicon wave heated and has attained a density of ∼ $2\times10^{18}$ $m^{-3}$. Figure 7 shows the axial profiles of electron temperature and density. The *d* value is the separation between the permanent magnet and the top flange (z = 0). The magnetic field in the source decreases as *d* is increased.

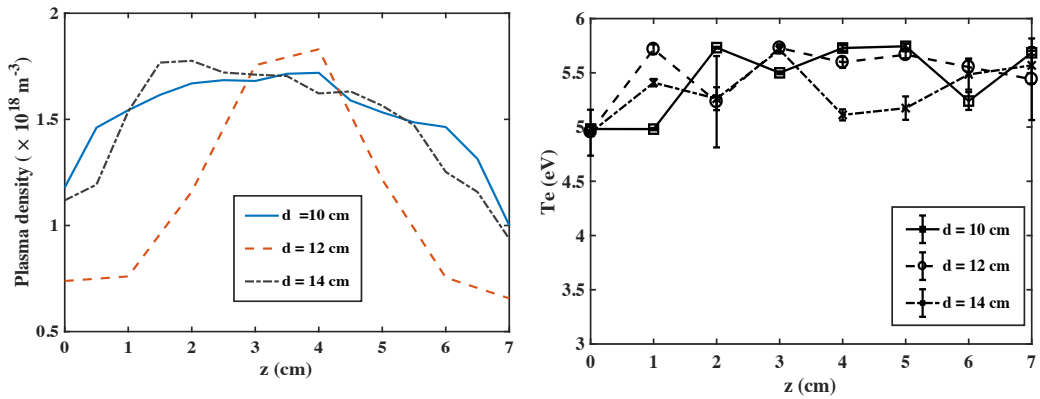

Figure 7. The axial density (a) and temperature (b) profile is shown. The maximum density obtained is ∼ $9\times10^{18} m^{-3}$ at 5 eV electron temperature with 800W power and 6 mtorr pressure.

For the three *d* values 10 cm, 12 cm and 14 cm, shown in the figure, the corresponding dc magnetic field values at z = 0 are ~ 40 G, ~ 55 G and 86 G respectively. For a fixed *d*, the value of the magnetic field decreases as we go downstream and is around 4 - 7 Gauss at z = 19 cm for the three cases considered. The location z = 19 cm has a diagnostic port for radial measurements in the expansion chamber and the $H^-$ ion measurements are carried out at z = 19 cm. This is discussed in the next section.

Figure 8 shows the radial profile of the axial component of the helicon wave field, $B_z$ measured by a B-dot probe. The radial profile shows that the $B_z$ is minimum at the centre and maximum approximately at $r = 2\ cm$, on both sides. This is a typical m = 1 mode helicon wave signature. Since $B_z \propto J_1(k_\perp r)$, figure 8 is fitted with the Bessel function to calculate the value of $k_\perp$ which comes out to be ~1.8 cm$^{-1}$.

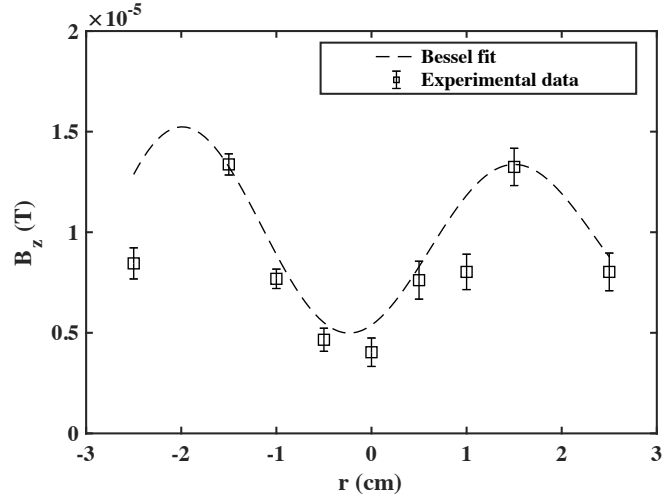

Figure 8. Radial field profile of $B_z$ measured by B-dot probe.

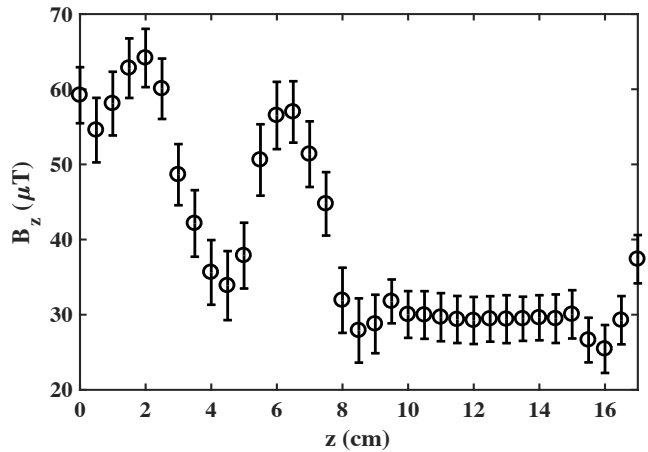

Figure 9. Typical axial wave field profile measured by B-dot probe.

Figure 9 shows an axial wave magnetic field profile. The axial component of the wave magnetic field ($B_z$) is measured using a B-dot probe calibrated at 13.56 MHz. The $B_z$ profile shows an axial propagation and damping. The component $B_z$ suddenly gets completely damped after z = 11 cm, which indicates a sudden wave absorption at that axial location. Further investigation is on-going to understand the phenomena and will be reported in future.

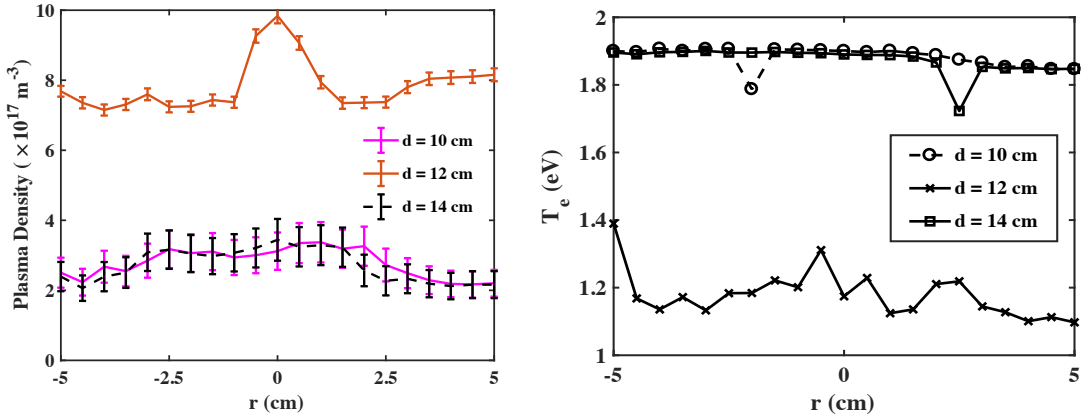

Figure 10. (a) Radial plasma density and (b) axial plasma temperature profile at z = 19 cm from the source in expansion region for different ring magnet positions with 800W RF power and 6mtorr pressure.

Fig. 10 shows the plasma density and temperature in the expansion region where $H^-$ ion density is measured. The parameter *d* is the separation between the top flange of the source and the centre of the ring magnet. The ring magnet separation at d = 12 cm from the driver back plate, is optimum as the plasma density is significantly higher and plasma temperature is significantly lower for that configuration, which are suitable for good negative ion yield.

We have measured the $H^-$ density in the downstream region using several diagnostic techniques to show that the HELEN configuration is suitable for negative hydrogen ion formation. The results are shown in the following section. It is to be noted that the observed $H^-$ density is lower than the electron or hydrogen ion ($H^+$) density. Moreover, in the presence of RF noise, extracting $H^-$ ion signal is difficult.

### b. Measurement of $H^-$ ion density

***OES method:*** The method of $H_\alpha/H_\beta$ line ratio is applied in HELEN-I for the measurement of negative ion density [7]. The line emission is measured with Ocean

Optics USB 4000 spectrometer. The spectrometer operates in the wavelength range of $\lambda = 200 - 1100$ nm with a spectral resolution of 2 nm. Emission light from the plasma is focussed to couple with an optical fibre using a lens and the light signal is transported to the spectrometer with the help of that optical fibre. The diameter of the collimating lens is 12 mm and focal length = 25 mm. A typical spectrum is shown in fig. 7 and such spectra are recorded for three different working pressures and for five different magnetic field configurations.

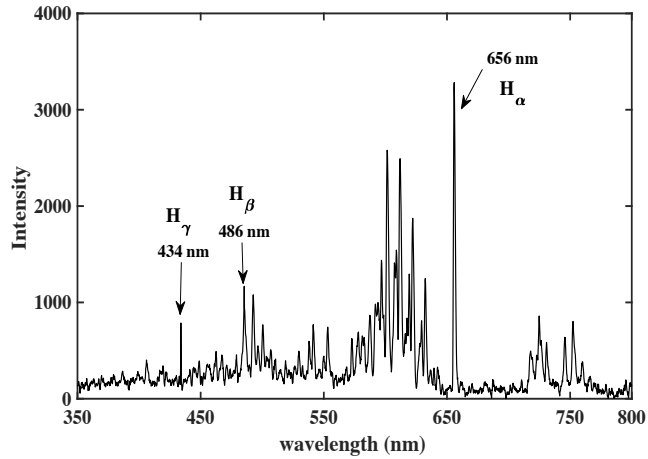

Fig. 11. OES spectrum showing different atomic and molecular lines

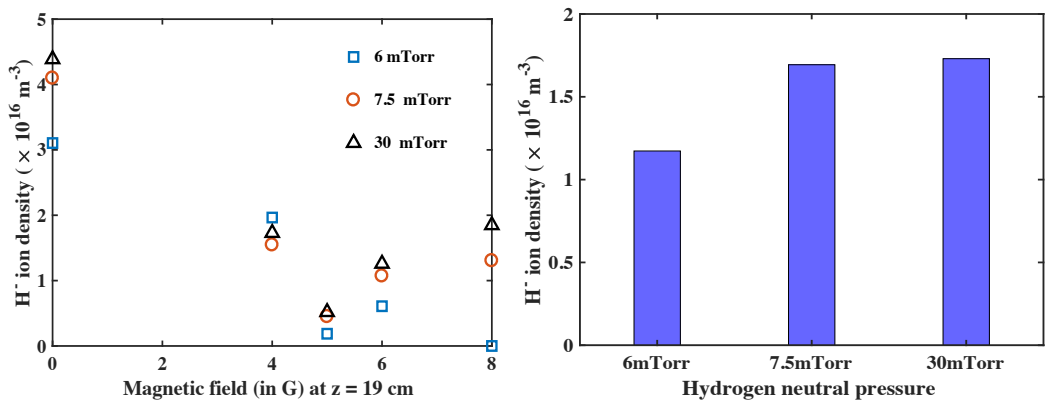

Figure 12. (a) $H^-$ density obtained from the OES measurement and its variation with the externally applied dc magnetic field due to placement of ring magnets at different axial positions; (b) shows the $H^-$ density for different operational pressure averaged over different magnetic field values at 800kW applied RF power .

The spectrum obtained from the OES is shown in figure 11. The $H_\alpha = 656\ nm$, $H_\beta = 486\ nm$ and $H_\gamma = 434\ nm$ lines are marked in the figure. These are the important emission lines for the calculation of $H^-$ ion density measurement as discussed in section III.2. The $H^-$ density is calculated using equations (1), (2) and (3) and the result is shown in figure 12. The values of effective emission rate coefficients are calculated from the Atomic Data and Analysis Structure (ADAS) system [25]. $H^-$ density close to $\sim 10^{16} \, m^{-3}$ is obtained from the line ratio measurements at 800W rf power. This measurement gives the $H^-$ density averaged over the plasma volume along the line of sight with ~ 40% error associated with the density values [7]. To verify the OES measurements a more sensitive technique CRDS is used. From equation (3) it can be seen that the $H^-$ density values measured by OES, depend inversely on the electron density. This is shown in figure 12.a where for the *0* gauss field the $H^-$ density is higher as the plasma density is lower for that case and decreases as the field is increased. This indicates that in presence of axial magnetic field energetic electrons are confined better which increases the electron-detachment destruction process of the H- ions. However, it is also true that axial field improves plasma density due to higher ionization by those confined energetic electrons.

*CRDS method*: The CRDS system attached to the experimental setup is shown schematically in figure 3. The preliminary results obtained from the two experiments are shown in figure 13 below. These values are obtained at 6mTorr neutral pressure and 5 Gauss field (ring magnet is placed 12 cm from the RF driver back plate) at the location (z = 19cm from back plate) where the CRDS data is obtained. A detailed description of the present CRDS system and parametric CRDS study of negative ion density variation is given in ref. 22 and 23 respectively.

It is observed that negative ion signals become detectable only after a certain power threshold (600 W) is crossed [23]. There are certain factors leading to this result. The plasma potential and electron temperatures are higher in the downstream region at lower applied rf power because the plasma is not yet in the helicon mode as can be seen from figure 5 in reference 13.

Typical Ring Down Time (RDT) plot of the CRDS experiment is shown in figure 13. The ring down time is obtained by fitting an exponential function on the acquired data. Equation (4) is used to calculate $H^-$ ion density for different cases using CRDS, where, time constant is $T_1$ is the reference (vacuum without plasma) decay time and $T_2$ is the decay time in the presence of the plasma.

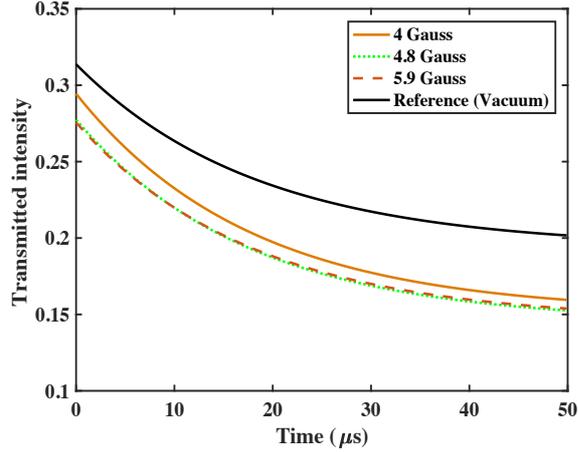

Figure 13. Typical exponential fits of the decay of different CRDS signals obtained by the photo detector are shown for different hydrogen plasma operation cases with slightly different magnetic field configurations.

Figure 14 shows a comparison between the $H^-$ density measurements done by CRDS for two different RF power cases. From fig. 12 and fig. 14, it is observed that OES gives slightly less $H^-$ density but of the same order, i.e. $\sim 10^{16}\ m^{-3}$. CRDS in HELEN-I setup is at z = 19 cm, which is at a distance of 12 cm from the driver mouth. The discrepancy is within the error bars. .

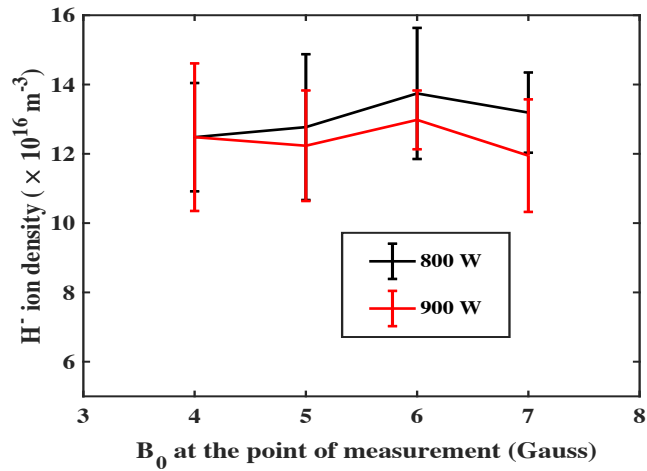

Figure 14. $H^-$ density obtained from CRDS at two different RF powers at 6 mTorr $H_2$ pressure for different magnetic field configuration due to different positions of the ring magnet

To understand the experimentally obtained values of negative ion density from both OES and CRDS, a simplified particle balance calculation has been carried out based on the formulation given in [9]. Considering steady state dominant production and destruction terms for $H^-$ ions, particle balance model can estimate the number density of $H^-$ available in the source. The main processes responsible for the formation of negative hydrogen ions are vibrational excitation of $H_2$ molecule having vibrational level $v" > 4$:

$$e + H_2(v') \rightarrow e + H_2(v"); \qquad v" > v'.$$

and subsequent Dissociative Attachment (DA) with low energy electrons. The $H^-$ destruction processes include Electron Detachment (ED), Mutual Neutralization (MN) and Associative Detachment (AD). The density of $H_2(v")$ is calculated by the relation given in [9], considering the plasma condition in the driver. In the calculation cross-sections for $v" > 4$ is considered since below that vibrational level, DA cross-section is significantly low [25]. In the present case, the observed driver plasma temperature is ~ 5 eV. For these parameters the number density of excited $H_2$ molecules is calculated by the following expression:

$$n_v = n_{H_2} \frac{(f_e \langle \sigma v \rangle)_{EV}}{[(f_e \langle \sigma v \rangle)_{DA} + (f_e \langle \sigma v \rangle)_{Dis} + (f_e \langle \sigma v \rangle)_{ion}] + \frac{1}{n_e \tau_v}} \qquad (5)$$

where, $n_v$ is the $H_2(v")$ density, $n_{H_2}$ is the $H_2$ density, $f_e$ is the fraction of electrons participating in a given reaction, EV in the subscript corresponds to the $H_2(v")$ formation process by collision of $H_2(v)$ with fast electrons. *Dis* and *Ion* are dissociation and ionization processes respectively, responsible for $H_2(v")$ destruction, $\tau_v$ is the residence time of the molecule, taken to be ~ 10 $\mu s$ considering molecules and neutrals at room temperature confined within the chamber walls. After $H_2(v")$ formation in the driver, they get transported to the expansion volume where plasma density and temperature are low and are conducive for $H^-$ production. The electron density in the expansion volume reduces and which leads to a reduction in the rate of formation of $H_2(v")$. But, the temperature in the downstream region is lower as compared to the driver region. Therefore, the probability for dissociative attachment is more and the probability for $H_2(v")$ destruction is less as compared to the driver region. DA of electrons is the primary process by which $H^-$ ions are formed. The final expression for $H^-$ density is [9]:

$$n_- = n_e \frac{-\left[\langle \sigma v \rangle_{MN} + \langle \sigma v \rangle_{ED} + \frac{1}{n_e \tau_-}\right] + \sqrt{\left[\langle \sigma v \rangle_{MN} + \langle \sigma v \rangle_{ED} + \frac{1}{n_e \tau_-}\right]^2 + 4\left[\langle \sigma v \rangle_{MN} \langle \sigma v \rangle_{DA} \frac{n_v}{n_e}\right]^2}}{2 \langle \sigma v \rangle_{MN}} \qquad (6)$$

The cross-sections for the electron-neutral collision and subsequent vibrational excitation of the $H_2$ molecules are taken from references [9], [26], [27] and [28]. The excitation of $H_2$ molecule occurs due to its collision with fast electrons with $T_e > 20\ eV$ [27]. In the present experiment we have a plasma with average electron temperature ~ 5 eV in the source and ~1eV in the expansion region. Assuming the electron distribution to be Maxwellian, we see that only the tail electrons in the distribution would participate in the vibrational excitation of the $H_2$ molecules. Since the electron population is assumed maxwellian, the number of electrons in an energy interval between $E$ and $E+dE$ can be calculated from the expression:

$$n_{dE} = n_e A_1 \int_E^{E+dE} E \exp\left(-\frac{1/2\ mv^2}{E_{th}}\right) dE \tag{7}$$

Here, $A_1$ is the normalization constant; $n_e$ is the total electron density, $E_{th}$ is the mean thermal energy of the particle distribution and $E = \frac{1}{2}mv^2$ is the kinetic energy of the particles. Since, different production and destruction processes in equation (5) require electrons of different energies present for example, the EV process requires electrons with energy exceeding 20 $eV$ and ionization requires $T_e > 13\ eV$.

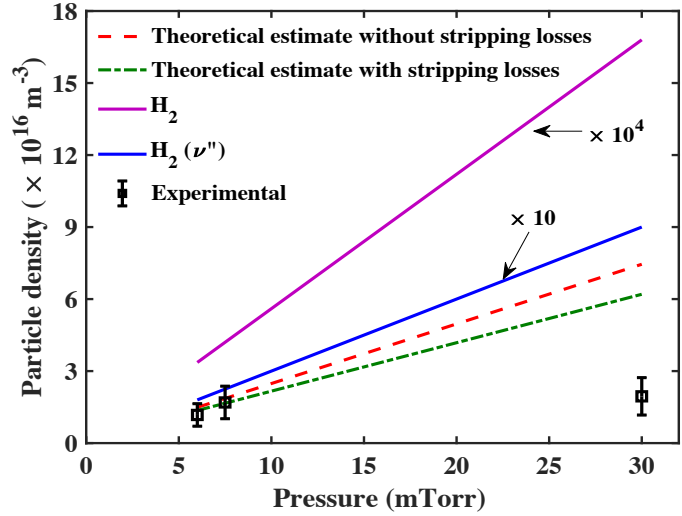

Figure 15. The theoretically obtained value for the $H^-$ density is plotted with the experimental values (same as figure 12.b). The $H_2$ molecule density and the density of vibrationally excited $H_2(v'')$ molecule are also shown to compare the destruction and generation processes.

Therefore, the population of electrons in an energy bin *dE* is calculated and used in equation (5) to obtain the excited population of $H_2(v'')$ molecules. This value is used in equation (6) to get the negative hydrogen density value. Here, $\tau_-$ is the confinement time for $H^-$ ions before escaping to the walls. Assuming the $H^-$ ion temperature to be ~ 0.1 eV, confinement time ~ 1 $\mu s$. The $H^-$ density thus calculated from equation (6) for different pressures is plotted with the experimentally obtained $H^-$ density value obtained from OES and CRDS in figure 15. The equations give a good estimate of the $H^-$ density, which is very close to the values obtained in experiment for lower pressure but deviates from the experiment at high pressure. The theoretically expected density value shows a linear trend since it depends linearly on the number of $H_2$ molecules present (equation 4). But since equations 4 and 5 only take into account the production and destruction mechanism at one particular location and not the transport of the species involved ($H^-, e, H_2(v")$ and $H^+$), they can only be used to get an estimate of the expected particle density. The plot of theoretically expected density of $H^-$ after including the stripping losses due to the collisions with energetic H atoms (associative detachment reaction) is close to the experimental value at lower pressures.

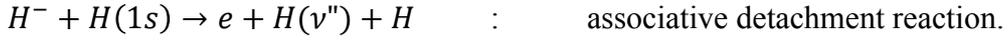

$H^- + H(1s) \rightarrow e + H(v") + H$     :     associative detachment reaction.

At low energy of collision, the reaction of $H^-$ with H atom is important and collision with $H_2$ molecules can be neglected [30]. $H^-$ ions are also lost due through diffusion to walls during transport in addition to the $H^-$ destruction processes described above. It can also be seen from figure 15 that the slope of the plot showing $H_2(v")$ density is higher than $H^-$ density plot, predicting that the destruction processes are dominant.

## V. Conclusion and Summary

A permanent ring magnet based hydrogen helicon plasma source has been developed. Performance characterization in hydrogen helicon plasma is reported in this manuscript. The plasma expands from driver region to expansion region due to the geometry as well as diverging field. This configuration is suitable for low electron temperatures and high plasma density in the downstream region of the plasma. The line integrated $H^-$ density is measured using two different diagnostic techniques namely Cavity Ring Down Spectroscopy and Optical Emission Spectroscopy. The two diagnostics used give the $H^-$ density in the order of $10^{16}\ m^{-3}$. A particle balance model based calculation also predicts similar density values considering experimental plasma conditions. The theoretically estimated value matches closely with the OES measurements at low pressures. But at high pressure, the stripping losses become prominent. The stripping losses are due to collisions of $H^-$ ions with $H$ atoms, $H_2$ molecules, $H^{3+}$ ions and electrons [31]. We have only considered the collisions with $H$ atoms and electrons here and we see that the measured value deviates from the theoretically estimated value. Also, due to the reasons mentioned in section III.2 at pressures higher than 7.5 mTorr, the results obtained from the line ratio method are expected to deviate from the actual value [7]. The production of $H^-$ ions in the system is through volume process, without Caesium (Cs) injection. We are able to obtain a good yield of $H^-$ ions even without the Cs injection or the transverse filter fields. This is due to the temperature gradient achieved due to geometrical and magnetic expansion of plasma in the expansion chamber. Injection of Cs vapour into the plasma volume converts the mode of plasma source operation from volume mode to surface mode, where $H^-$ ions are mainly formed on low work-function surfaces instead of collisional based volume process. In future, Cs injection has to be established in HELEN plasma source. In addition, research with filter fields and low temperature electron sources are planned in HELEN to improve the $H^-$ yield. A laser photo detachment based $H^-$ ion density measurement using Langmuir probe is under development for local density measurements to generate the $H^-$ ion density profile.